\documentclass{article}
\usepackage{spconf,amsmath,graphicx}
\usepackage{booktabs}
\usepackage{multirow}
\usepackage{array}
\usepackage{color}
\usepackage{url}
\usepackage{subfigure}
\usepackage{amssymb}

\newcommand{\etal}{\textit{et al}.}

\def\w{4cm}
\def\h{3cm}
\title{Nonlinear Prediction of Multidimensional Signals via Deep Regression \\
		 with Applications to Image Coding}
\name{
	Xi Zhang $^{\star}$ \quad Xiaolin Wu $^{\star \dagger}$
}
\address{
	$^{\star}$ Department of Electronic Engineering, Shanghai Jiao Tong University \\
	$^{\dagger}$ Department of Electrical \& Computer Engineering, McMaster University
}

\begin{document}
%\ninept
%
\maketitle
\begin{abstract}
Deep convolutional neural networks (DCNN) have enjoyed great successes in many signal processing applications because they can learn complex, non-linear causal relationships from input to output.  In this light, DCNNs are well suited for the task of sequential prediction of multidimensional signals, such as images, and have the potential of improving the performance of traditional linear predictors.  In this research we investigate how far DCNNs can push the envelop in terms of prediction precision.  We propose, in a case study, a two-stage deep regression DCNN framework for nonlinear prediction of two-dimensional image signals.  In the first-stage regression, the proposed deep prediction network (PredNet) takes the causal context as input and emits a prediction of the present pixel.  Three PredNets are trained with the regression objectives of minimizing $\ell_1$, $\ell_2$ and $\ell_\infty$ norms of prediction residuals, respectively.
The second-stage regression combines the outputs of the three PredNets to generate an even more precise and robust prediction.
The proposed deep regression model is applied to lossless predictive image coding, and it outperforms the state-of-the-art linear predictors by appreciable margin.
\end{abstract}
\section{Introduction}
% ~\cite{rissanen1981, rissanen1983, rissanen1984}

%Generalizing universal prediction, modeling and coding to multidimensional signals (e.g., images and videos) has great engineering value, because it offers a more principled approach for many tasks of multidimensional signal processing: compression, denoising, restoration, classification, etc.
%
%In the field of traditional predictive coding, a multidimensional signal has to be sequentialized to fit into Rissanen's framework of universal sequential prediction. For example, a 2-D image signal $x(i,j)$ of N pixels has to be mapped into a sequence $x_1, x_2, ..., x_N$, before algorithm $Context$~\cite{rissanen1983}, which was devised for sequential prediction of 1-D random process, can be applied to $x(i,j)$. The algorithm will make different predictions under different sequentializations of $x(i,j)$. A predetermined, signal-independent sequentialization of $x(i,j)$ can obscure statistical dependencies among the pixels and degrade the prediction performance. Adding to the complexity of the problem is that the sequentialization should be allowed to change in pixel position $(i,j)$ because the image signal $x(i,j)$ is often nonstationary.

Sequential prediction of signals plays important roles in many applications, ranging from economics to image/video processing.  Practically, all existing predictors used in image/video processing and computer vision are linear.  This linearity is not due to the nature of the underlying physical problems; instead, it is only the result of operational expediency.  Optimal design of linear predictors is computationally intractable.
Linear prediction is effective to decorrelate stationary Gaussian random process, and is widely used in predictive coding of multidimensional signals. The classical linear predictors for image coding can be found in~\cite{wu1998,li2001,takeda2006,kervrann2008,memon2000,akimov2007}.

Even living with the limitation of linear predictors, there is another difficulty hindering the optimal design of linear predictors of image signals; that is, the choice of causal context for predicting the current pixel.  The standard practice is to use the template that contains the $K$ closest known pixels to the current pixel.
% as shown in Fig.~\ref{fig:causal_demo}.
The order $K$ of the prediction model is fixed throughout the sequential prediction process and chosen empirically. The 2-D prediction context is simply a rectangular causal region of size $K$ that is centered at the next pixel $x_i$. Justifying this design is the assumption that the correlation between two samples increases as they get closer to each other in space/time. Although the assumption might be true for many 1-D signals (e.g., ECG, audio), it does not hold for multidimensional signals as sample dependencies in natural signals are anisotropic in general.  As such, a signal-independent prediction context must be suboptimal because it includes irrelevant past samples and misses relevant ones.

%The drawbacks of predetermined predictor support were noticed by some researchers. For example, Takeda~\etal~\cite{takeda2006} suggested to use an ellipse shaped support, with the major axis aligned with the direction of local edge. Kervrann and Boulanger~\cite{kervrann2008} proposed a model selector for choosing the size of square support according to local image statistics. In~\cite{memon2000}, Memon~\etal studied the optimal order of scanning pixels in the perspective of context modeling for lossless image coding. They analyzed the influence of the pixel scan order on the context support but did not discuss the design of optimal sequential predictors. In their analysis, the image was modeled as an isotropic Gaussian random field, and accordingly the context support was made of the causal pixels within a fixed radius of the current pixel, which are too restrictive as we argued previously. In~\cite{akimov2007}, Akimov~\etal proposed several pixel ordering schemes for context tree modeling in arithmetic coding of color map images.

Wu~\etal~\cite{MDL-PAR} proposed an adaptive, piecewise autoregressive (PAR) prediction model for multidimensional signals.  It uses the correlation instead of Euclidean distance between the past sample $x_{i-t}$ and the current sample $x_i$ to sequentialize past samples to form spatially nested causal prediction contexts for different orders of the PAR model.
For each $x_i$, the order of the PAR model is determined in a criterion of minimum description length (MDL).  To estimate the PAR model parameters for $x_i$, the authors also developed a technique to choose a causal training set of past samples and the associated prediction context. The MDL model is optimally designed on a sample-by-sample basis, and it beats all of its predecessors by achieving the lowest entropy of prediction residuals up to now.  However, the MDL optimization process proposed in~\cite{MDL-PAR} has a prohibitively high computational complexity, requiring 8 hours to perform sequential prediction of a 512$\times$512 image.

This research is inspired by great successes of deep learning in various signal processing applications, aiming to use the new tool to improve the performance of existing predictors for multidimensional signals.  Our goal is well within reach because deep convolutional neural networks can learn complex, non-linear causal relationships, provided that a large amount of paired input and output data is available.  In addition to breaking the linearity limit, a DCNN prediction model also circumvents the difficulty of finding a suitable prediction context because it can, via the training process with a sparsity constraint, discover effective features that contribute to accurate prediction.

Operationally, a deep learning based predictor also has advantage.  Although the training of the DCNN prediction model is computationally expensive, it is only an off-line process.  At the on-line inference stage, the new method runs faster than the state-of-the-art method of adaptive MDL predictor.

In this paper, the technical developments are presented mostly around 2D image signals.
However, the ideas and results can be easily extended to signals of other dimensions.
We propose a two-stage deep regression DCNN for nonlinear prediction of 2D signals.
In the first-stage regression, the prediction network (PredNet), consisting of a convolutional module and a regression module, is designed to take the causal context as input and output the prediction of the current pixel.
Three PredNets are trained with the different regression objectives of minimizing $\ell_1$, $\ell_2$ and $\ell_\infty$ of prediction residuals, respectively.
In the second-stage regression, called refinement regression, the different predictions from the three trained PredNets are fed into a new regression network to generate a more precise and robust prediction for the current pixel.
To validate the effectiveness of the proposed two-stage deep regression DCNN, we apply it to lossless image coding and evaluate the self-entropy of the prediction residuals.  The new deep learning prediction method achieves the lowest entropy of the prediction residuals, among all predictors that have been published till present.

% \section{Related Works}

\section{Sequential Prediction via Deep Regression}

\subsection{Problem Formulation}
\begin{figure}
	\centering
	\includegraphics[width=6cm]{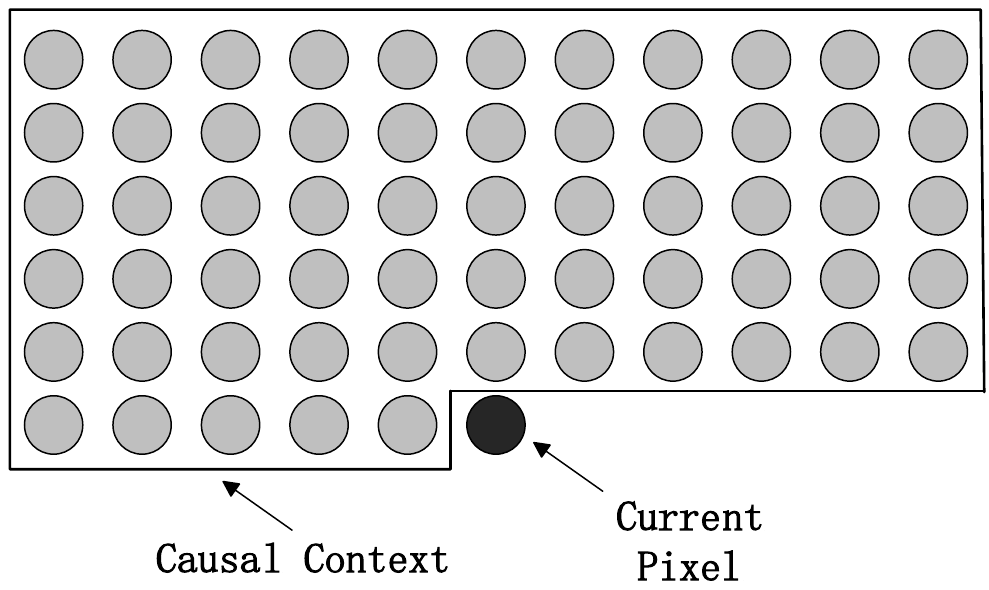}
	\caption{The illustration of the 2-D predictor and the corresponding causal context. One circle represents one pixel in the image. }
	\label{fig:causal}
\end{figure}

For an image signal modeled as a 2D Markov field, the sequential prediction of the current pixel $x$ is made in a suitable causal neighborhood, that is:
\begin{align}
	\hat{x} = F(C(x))
\end{align}
where $C(x)$ is a causal context consisting of past pixels that have effects on $x$; a simple prediction context of nearest neighbors is illustrated in Fig~\ref{fig:causal}.  In what follows, we investigate how the predictor $F$ can be realized by a prediction neural network model (PredNet) of deep learning.

Given a set $S$ of training samples $\{ x_i ; C(x_i) \}$, the PredNet can be optimized by solving the following minimization problem:
\begin{align}
	F = \arg \min_F E_{x\in S} \| F(C(x)) - x \|_\ell
\end{align}
where $E$ represents the expectation over the training set $S$.

\subsection{Minimum-entropy prediction}
For the training of the DCCN prediction model $F$, any $\ell$-norm of the prediction residuals can be used in the objective function.  Different $\ell$-norms can be chosen to serve different design purposes.  In lossless image compression, for instance, the ultimate goal is to minimize the entropy of the prediction residuals.  Designing minimum entropy predictor, despite its practical values in data compression, has been hardly studied, apparently because of the difficulty of the problem.  The only known work is a linear minimum entropy predictor by Wang and Wu \cite{wang2007}, which is computed by convex or quasiconvex programming.  Now with the new tool of DCCNs, we embark on designing non-linear minimum-entropy predictors, which has remained to be a hard nut to crack thus far.

For most natural images prediction residuals obey a Laplacian distribution~\cite{laplacian}; hence, minimizing $\ell_1$-norm is equivalent to minimizing the entropy of prediction residuals.
% In order to compare the differences between the prediction networks optimized by different $\ell$-norms, we select $\ell=1, 2, \infty$ in our experiments.

For the sake of completeness, besides the $\ell_1$-norm selected as a proxy for minimizing the entropy, we also design non-linear DCNN predictors of minimum $\ell_2$ and $\ell_\infty$ norms. The three prediction networks (PredNets) trained with the criteria of minimum $\ell_1$, $\ell_2$ and $\ell_\infty$ are called PredNet-$\ell_1$, PredNet-$\ell_2$ and PredNet-$\ell_\infty$, respectively.

\subsection{Network Architecture}
\begin{figure*}[h]
	\centering
	\includegraphics[width=16cm]{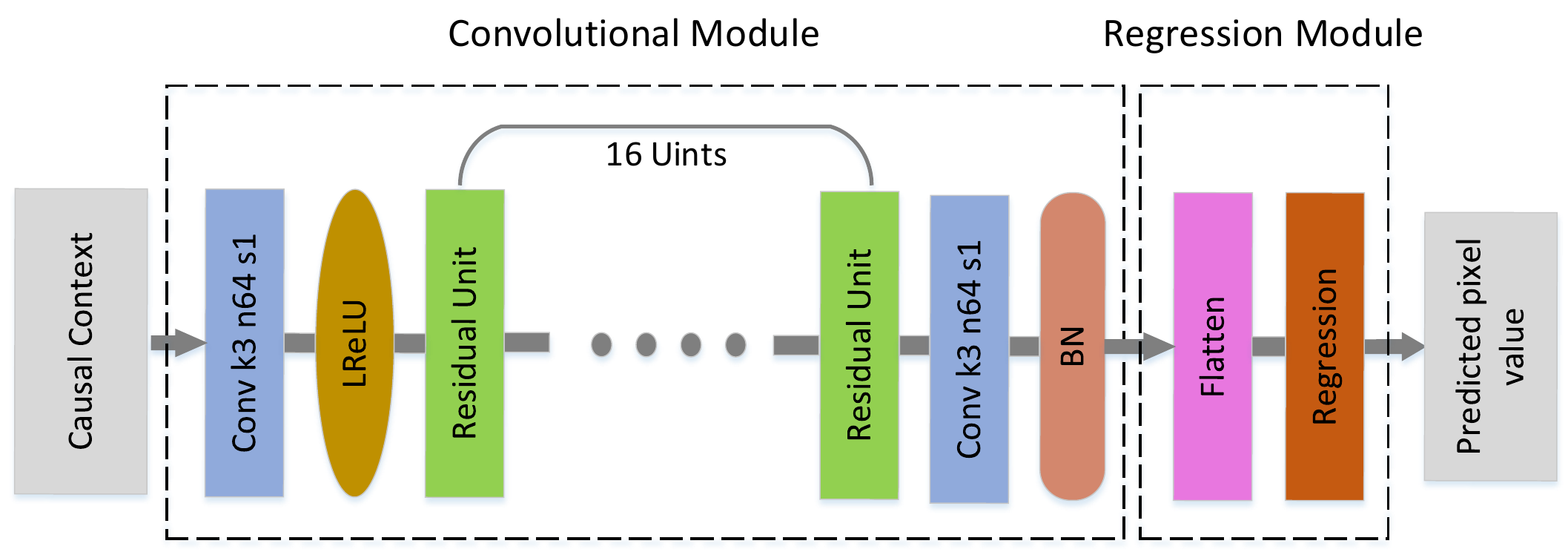}
	\caption{The architecture of the deep prediction network (PredNet).}
	\label{fig:predNet}
\end{figure*}

\begin{figure}[h]
	\centering
	\includegraphics[width=6cm]{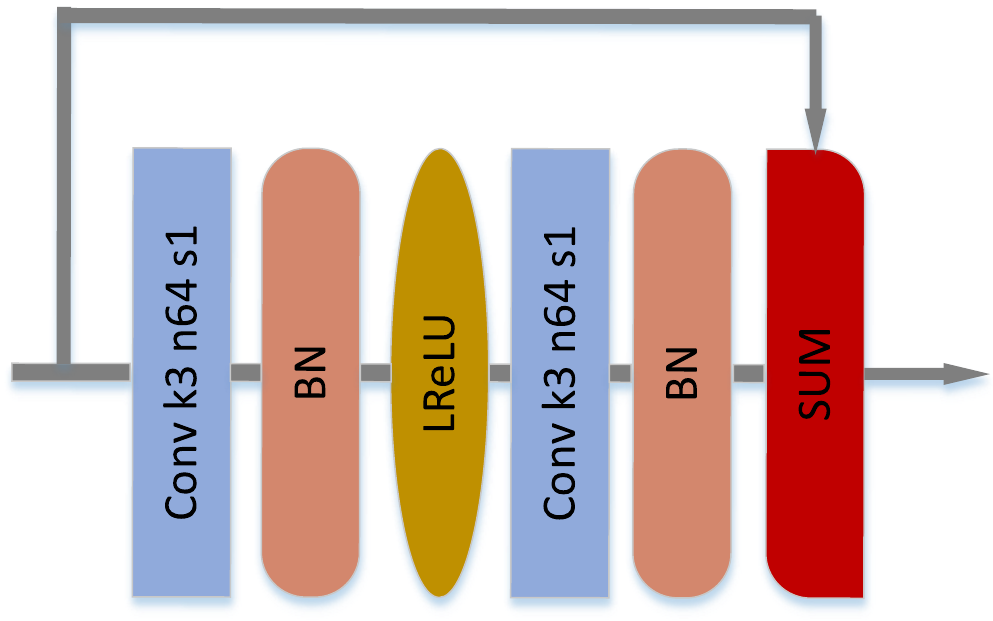}
	\caption{The detailed configurations of the residual unit.}
	\label{fig:res_unit}
\end{figure}

The proposed PredNet consists of a convolution module and a regression module.
The convolutonal module is designed to extract features that contribute to the prediction from the causal context, and the regression module applies the regression on these extracted features to emit the prediction.
As illustrated in the Fig.~\ref{fig:predNet} and Fig.~\ref{fig:res_unit},
The convolutional module contains 16 residual units. Each unit consists of two convolutional layers, respectively followed by a batch-normalization layer and a LeakyReLU activation layer. For LeakyReLU, the slope of the leak is set to 0.2.
The regression module contains a flatten layer and a fully-connected regression layer with linear activation.


\subsection{Sparse Regularization}
If the image is modeled as a Markov random field, then the input of the PredNet, i.e., the causal prediction context, should be sufficiently large to contain all the pixels that influence the current pixel, but at the same time it may include some irrelevant pixels.
We rely on the deep regression of DCNN to discover useful features that contribute to the prediction and in the process discard the irrelevant pixels.  In the principle of MDL, or to reduce the risk of overfitting, we promote a compact DCNN prediction model by requiring the coefficients of the regression module to be sparse.  By noting that the weights $w_r$ of the regression layer behave like a selection function in pixel domain, we include a model cost regularization term $R(w_r)$ into the objective function for training the prediction network $F$:
\begin{align}
	F = \arg \min_F
		\bigg\{
				E_{x\in S} \| F(C(x)) - x \|_\ell
						+ \lambda R(w_{r})
		\bigg\}
\end{align}
where the scalar $\lambda$ is a Lagrangian multiplier.  Here we adopt the most common form of sparse regularization in neural networks, the $\ell_1$-norm of neuron weights, namely,
\begin{align}
	R(w_r) = \| w_r \|_1
\end{align}

\subsection{Refinement Regression}
In the first-stage regression, three deep prediction networks have been trained for minimizing different norms of prediction residuals.
For pixel $x$, the three predictions optimized in the different criteria are denoted by $\hat{x}_{\ell_1}$, $\hat{x}_{\ell_2}$ and $\hat{x}_{\ell_\infty}$, respectively.
The goal of the second-stage regression is to train a refined DCNN prediction model that takes $\hat{x}_{\ell_1}$, $\hat{x}_{\ell_2}$, $\hat{x}_{\ell_\infty}$ as input and emits an improved prediction $\tilde{x}$.
Using another training set $S^{'}$ (different from $S$ to prevent overfitting), the refined regression network $\mathbb{F}$ can be trained by minimizing the following cost function:
\begin{align}
	\mathbb{F} = \arg \min_\mathbb{F} E_{x\in S^{'}}
		\|
		\mathbb{F}(\hat{x}_{\ell_1}, \hat{x}_{\ell_2}, \hat{x}_{\ell_\infty})
		- x
		\|_l
\end{align}
In the interest of lossless image compression, our goal is to design a DCNN minimum-entropy predictor, thus the $\ell_1$-norm of prediction residuals is to be minimized in the training of the refined regression network PredNet-R.

\begin{figure*}[!h]
\centering
\subfigure[Image]{
\includegraphics[width=\w, height=\h]{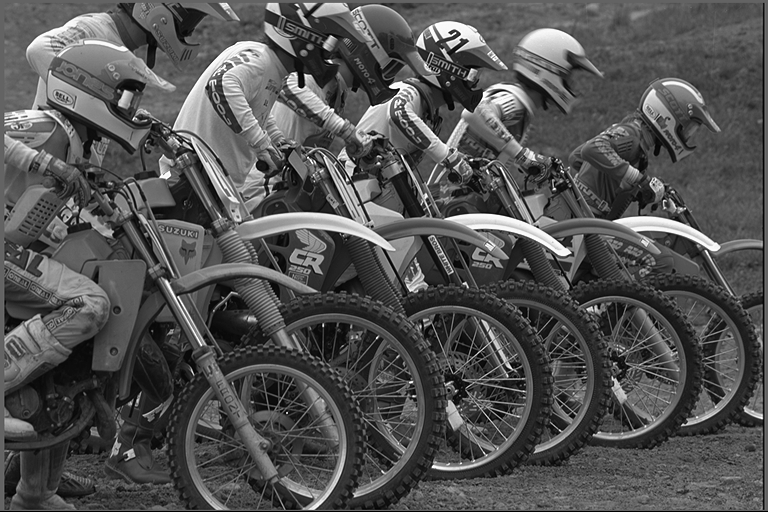}}
\subfigure[GAP]{
\includegraphics[width=\w, height=\h]{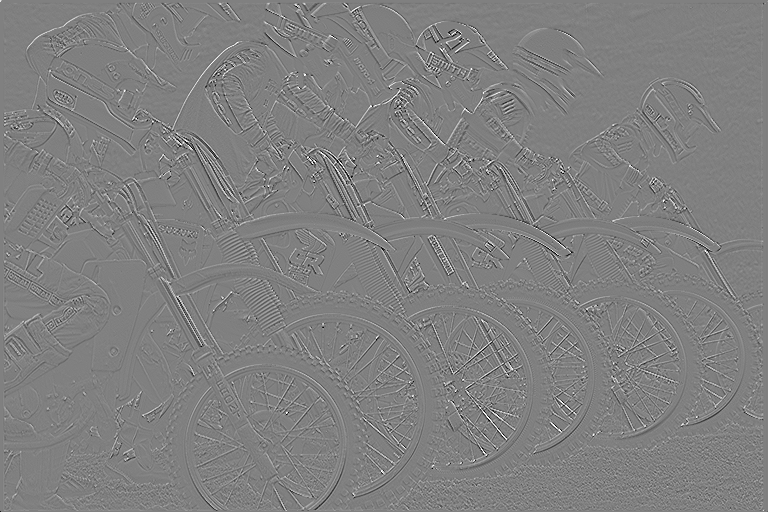}}
\subfigure[MDL-PAR]{
\includegraphics[width=\w, height=\h]{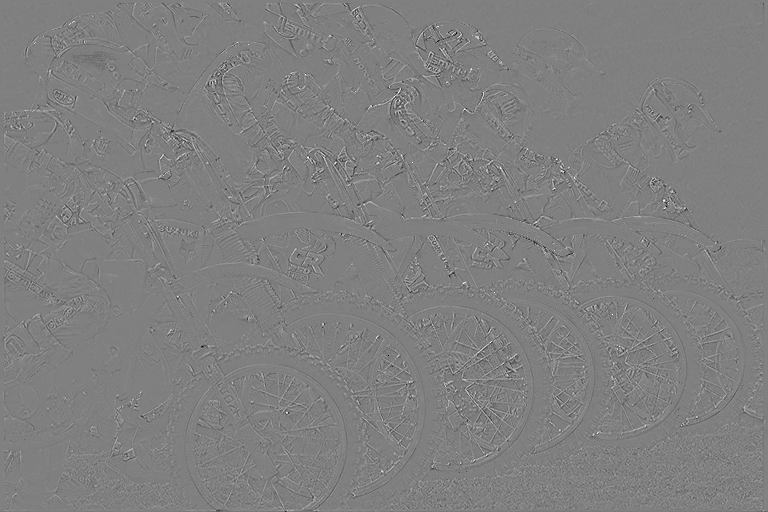}}
\subfigure[PredNet-R]{
\includegraphics[width=\w, height=\h]{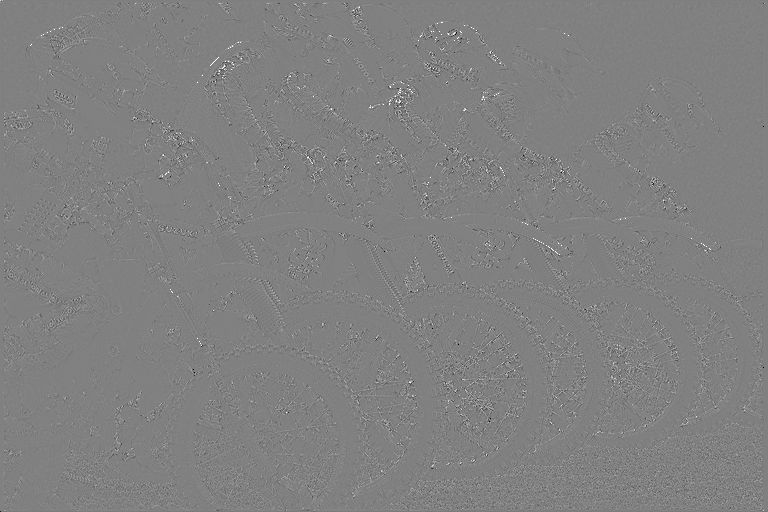}}
\caption{Residual images for `motocross bikes' from the Kodak image dataset.}
\label{fig:bikes}
\end{figure*}
\begin{figure*}[!h]
\centering
\subfigure[Image]{
\includegraphics[width=\w, height=\h]{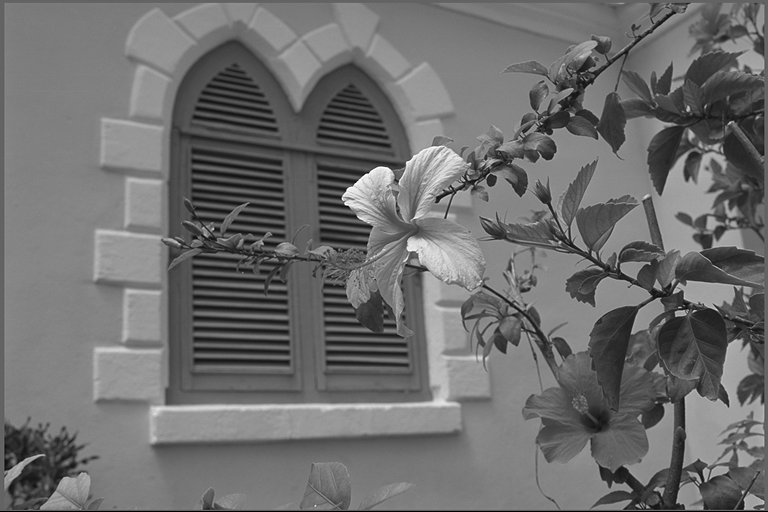}}
\subfigure[GAP]{
\includegraphics[width=\w, height=\h]{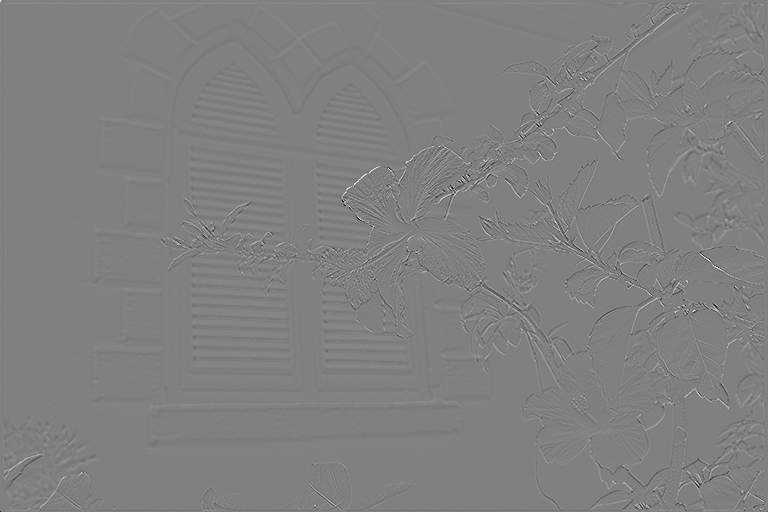}}
\subfigure[MDL-PAR]{
\includegraphics[width=\w, height=\h]{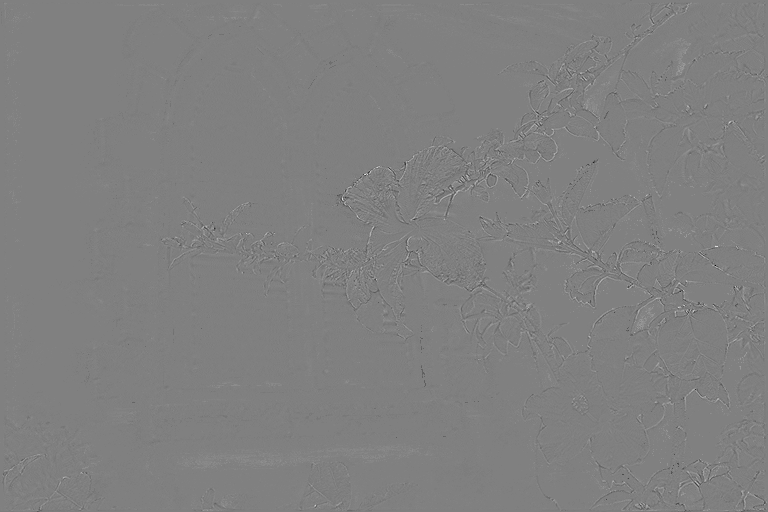}}
\subfigure[PredNet-R]{
\includegraphics[width=\w, height=\h]{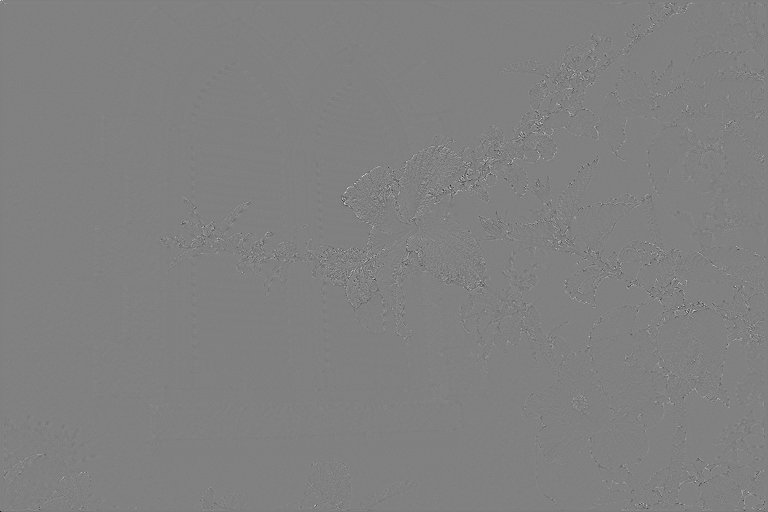}}
\caption{Residual images for `shuttered windows' from the Kodak image dataset.}
\label{fig:windows}
\end{figure*}

\section{Experiments}
To validate the effectiveness of the proposed deep prediction network (PredNet), we compare the prediction residuals on four measures ($\ell_1$, $\ell_2$, $\ell_\infty$ and the self-entropy), with the state-of-the-art predictor MDL-PAR~\cite{MDL-PAR}, and the gradient adaptive prediction (GAP) used in the well-known lossless compression algorithm CALIC \cite{CALIC}.
For training, we collect thousands of 2K-resolution high-quality images from the three public datasets: DIV2K~\cite{DIV2K}, CLIC~\cite{CLIC} and Flickr2K~\cite{Flickr2K}, then randomly extract millions of patches from these images as the training set.
Test images used in our experiments are from the Kodak lossless image dataset~\cite{kodak}.

In order to facilitate the computation, we use $\ell_8$-norm as an alternative to the $\ell_\infty$-norm in training PreNet-$\ell_\infty$.
The hyper-parameters used for training the PredNets are listed as following: size of causal context is $21 \times 21$; learning rate is fixed to $10^{-4}$; weighting coefficient $\lambda$ is set to $0.2$; the parameters of Adam optimizer is $\beta_1=0.9, \beta_2=0.99, \epsilon=10^{-8}$.

\begin{table}[!ht]
\centering
\caption{Performance comparisons with the state-of-the-art predictors. \textcolor{blue}{Blue numbers} indicate the best performances achieved in the first-stage regression; \textcolor{red}{Red numbers} indicate the best performances after the second-stage regression.}
\centering
\label{tab:self-ent}
\renewcommand\arraystretch{1.3}
\vskip 0.2cm
\begin{tabular}{c|p{0.8cm}<{\centering}p{0.8cm}<{\centering}p{0.8cm}<{\centering}p{0.8cm}<{\centering}p{0.8cm}<{\centering}}
\hline
\multirow{2}{*}{Predictors} & \multicolumn{4}{c}{Measures} \\
	& $\ell_1$ & $\ell_2$ & $\ell_\infty$ & entropy & $\rho_{max}$ \\
\hline
GAP  & 5.68 & 10.36 & 188.25 & 4.69 & 0.27 \\
MDL-PAR  & 5.34 & 9.65 & 181 & 4.40 & 0.24\\
\hline
PredNet-$\ell_1$ & \textcolor{blue}{4.51} & 9.40 & 241.29 & \textcolor{blue}{4.32} & \textcolor{blue}{0.20} \\
PredNet-$\ell_2$ & 4.62 & \textcolor{blue}{8.32} & 228.20 & 4.38 & 0.21 \\
PredNet-$\ell_\infty$ & 5.65 & {8.50} & \textcolor{blue}{160.29} & 4.58 & 0.24 \\
\hline
PredNet-R & \textcolor{red}{4.48} & 8.82 & 230.12 & \textcolor{red}{4.25} & \textcolor{red}{0.18}\\
\hline
\end{tabular}
\end{table}

The performance comparisons with the state-of-the-art predictors are listed in Table~\ref{tab:self-ent}.
As illustrated in the table, in the first-stage regression,  the DCNN predictors PredNet-$\ell_1$, PredNet-$\ell_2$ and PredNet-$\ell_\infty$ outperform the state-of-the-art predictor MDL-PAR in their respective error criterion.
%optimization objectives.
PredNet-$\ell_1$ not only has the smallest $\ell_1$-norm, it also has the lowest entropy of the prediction residuals.  This result indicates that the prediction residuals indeed obey the Laplacian distribution, hence minimizing $\ell_1$-norm is equivalent to minimizing the entropy of prediction residuals.

In the refinement regression, the DCNN predictor PredNet-R further reduces the $\ell_1$ error and the entropy of the prediction residual, by combining the different predictions in the first-stage regression.  PredNet-R exhibits the power and advantages of deep learning in multidimensional signal prediction over traditional methods by breaking the record of achievable lowest entropy held by the MDL-PAR predictor.  This achievement is remarkable considering the extremely high complexity of the MDL-PAR predictor that needs to solve one optimization problem per pixel.  As a result, it requires hours to preform sequential prediction of a 512$\times$512 image.  In contrast, the proposed deep learning predictors are designed off line, and they run much faster than the MDL-PAR predictor at the time of inference, taking 30 seconds per 512$\times$512 image.

In addition to the entropy, the performance of an image predictor can be measured by lack of correlation between the prediction residual and the original image signal. We compute the local maximum correlations (denoted by $\rho_{max}$) between the prediction residuals and the input image for the predictors in the comparison group, and include the results in Table~\ref{tab:self-ent}.  
The local maximum correlation refers to the largest of the correlation coefficients between the corresponding patches extracted from the prediction residuals and the original image.

The superiority of PredNet-R can be visualized by the absence of image structures in the residual image.  Figs.~\ref{fig:bikes} and \ref{fig:windows} are sample residual images of GAP, MDL-PAR and PredNet-R. It is evident that the residual images of PredNet-R contain the least amount of visible signal structures.
%indicating that the proposed deep regression framework is effective in image prediction.

\section{Conclusions}
In this work, DCNNs establish new performance records in sequential prediction of image signals.  The proposed deep learning signal prediction models may find applications in signal compression, denoising and analysis.

%We propose a two-stage deep regression DCNN framework for nonlinear prediction of two-dimensional image signals. The experiments demonstrate that the proposed PredNets are effective and robust.

% \vfill\pagebreak
\bibliographystyle{IEEEbib}
\bibliography{refs}

\end{document}